\documentclass[cits]{PoS}\usepackage{amsmath,cite,color}

\title{Instantaneous Bethe--Salpeter Look at the Lightest
Pseudoscalar Mesons}\ShortTitle{Instantaneous Bethe--Salpeter Look
at the Lightest Pseudoscalar Mesons}\author{\speaker{Wolfgang
Lucha}\\Institute for High Energy Physics, Austrian Academy of
Sciences, Nikolsdorfergasse 18, A-1050 Vienna, Austria\\E-mail:
\email{Wolfgang.Lucha@oeaw.ac.at}}\author{Franz F.~Sch\"oberl\\
Faculty of Physics, University of Vienna, Boltzmanngasse 5, A-1090
Vienna, Austria\\E-mail: \email{franz.schoeberl@univie.ac.at}}

\abstract{Within our description of Goldstone-type pseudoscalar
mesons as almost massless bound states of quark and antiquark by a
three-dimensional bound-state equation of Bethe--Salpeter origin,
taking into account the pointwise behaviour of the full
light-quark propagators enables to characterize the effective
interquark interactions more precisely than earlier studies
exploiting just specific~aspects.}

\FullConference{XVII International Conference on Hadron
Spectroscopy and Structure --- Hadron2017\\25--29 September,
2017\\University of Salamanca, Salamanca, Spain}\begin{document}

\section{Inversion Starting Point: an Instantaneous Bethe--Salpeter
Bound-State Equation}In the spectrum of hadron states, the
lightest pseudoscalar mesons, the pions and kaons, occupy a very
isolated position: On the one hand, they must be viewed as bound
states of a quark--antiquark pair. On the other hand, they
represent (nearly massless) Goldstone bosons related to the
dynamical chiral symmetry breakdown of quantum chromodynamics
(QCD), owing their nonzero masses only to some additional,
explicit breaking of the chiral symmetries. Needless to say, both
of these~aspects should be incorporated in any reasonable
description of such mesons within the framework of QCD.

The homogeneous Bethe--Salpeter equation constitutes a
Poincar\'e-covariant framework for the description of bound states
within quantum field theory. Problems due to the occurrence of
timelike excitations, inherent to fully relativistic formalisms,
may be evaded by relying on three-dimensional reductions
accomplished, for instance, by assuming all the effective
interactions experienced by the bound-state constituents to be, in
the center-of-momentum frame of the bound states, instantaneous.
Information on these effective interactions, extracted in form of
central potentials $V(r),$ $r\equiv|\mathbf{x}|$, may be deduced
by inversion of one's bound-state equation \cite{WL13}. Recently,
we did this for the presumably simplest reduction of this kind
\cite{WL-I}, the Salpeter equation \cite{SE}, and a
straightforward generalization of the latter \cite{WL-N}, designed
to incorporate more of the Bethe--Salpeter formalism's
relativistic nature \cite{WL05}.

In the present context, inversion simply means the reconstruction
of the effective interactions in one's bound-state equation from
available knowledge about solutions to this very equation. For
ease of presentation, let us impose flavour symmetry, by assuming
our two bound-state constituents~to be some quark and the
corresponding antiquark. Information about this required input to
the envisaged inversion process can be harvested by exploiting the
fact that the chiral symmetries of QCD provide, in the chiral
limit, an identity \cite{MRT} relating the quark propagator and
the Bethe--Salpeter solution for a massless pseudoscalar meson in
its center-of-momentum frame. If one's formalism strictly respects
Poincar\'e covariance, the full propagator $S(p)$ of a fermion of
four-momentum $p$ is totally defined by two Lorentz-scalar
functions, this fermion's mass $M(p^2)$ and wave-function
renormalization~$Z(p^2)$:$$S(p)=\frac{{\rm
i}\,Z(p^2)}{\not\!\!p-M(p^2)+{\rm i}\,\varepsilon}\ ,\qquad
\not\!\!p\equiv p^\mu\,\gamma_\mu\ ,\qquad\varepsilon\downarrow0\
.$$The semirelativistic bound-state equation constructed in
Ref.~\cite{WL05} as an instantaneous approximation to the
Bethe--Salpeter formalism poses an implicit eigenvalue problem,
with the bound-state masses $\widehat M$ as eigenvalues. Its
solutions, the Salpeter amplitudes $\phi(\mathbf{p})$, encode the
distribution of the relative three-momenta $\mathbf{p}$ of the
involved bound-state constituents. For \emph{pseudoscalar\/} bound
states of spin-$\frac{1}{2}$ fermions, the Salpeter amplitude
$\phi(\mathbf{p})$ involves only two independent component
functions $\varphi_{1,2}(\mathbf{p})$:
$$\phi(\mathbf{p})=\left[\varphi_1(\mathbf{p})\,
\frac{\gamma_0\,[\mathbf{\gamma}\cdot\mathbf{p}+M(p^2)]}
{\sqrt{\mathbf{p}^2+M^2(\mathbf{p}^2)}}
+\varphi_2(\mathbf{p})\right]\gamma_5\ .$$Assuming for the
effective interactions Fierz and spherical symmetry, our
bound-state equation~gets recast to a coupled set of one integral
and one algebraic relation \cite{WL07} for the radial functions
$\varphi_{1,2}(|\mathbf{p}|)$:\begin{align}
&2\,\sqrt{\mathbf{p}^2+M^2(\mathbf{p}^2)}\,\varphi_2(|\mathbf{p}|)
+\frac{4}{\pi}\,\frac{Z^2(\mathbf{p}^2)}{|\mathbf{p}|}
\int\limits_0^\infty{\rm d}|\mathbf{q}|\,|\mathbf{q}|
\int\limits_0^\infty{\rm d}r\sin(|\mathbf{p}|\,r)
\sin(|\mathbf{q}|\,r)\,V(r)\,\varphi_2(|\mathbf{q}|)=\widehat
M\,\varphi_1(|\mathbf{p}|)\ ,\nonumber\\[1ex]
&2\,\sqrt{\mathbf{p}^2+M^2(\mathbf{p}^2)}\,\varphi_1(|\mathbf{p}|)
=\widehat M\,\varphi_2(|\mathbf{p}|)\ .\label{GIBSE}\end{align}

\section{Inversion: Confining Interquark Potential Compatible with
Goldstone's Theorem}
\begin{figure}[h]\begin{center}\begin{tabular}{cc}
\includegraphics[scale=1.51645]{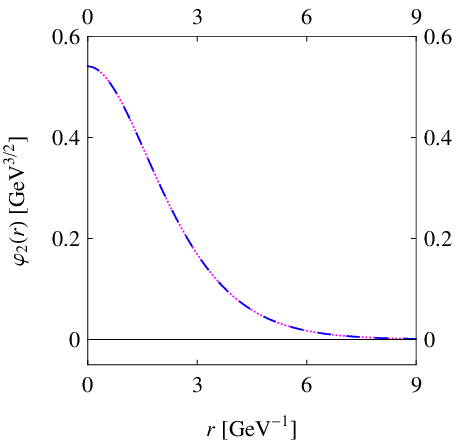}\quad&\quad
\includegraphics[scale=1.51645]{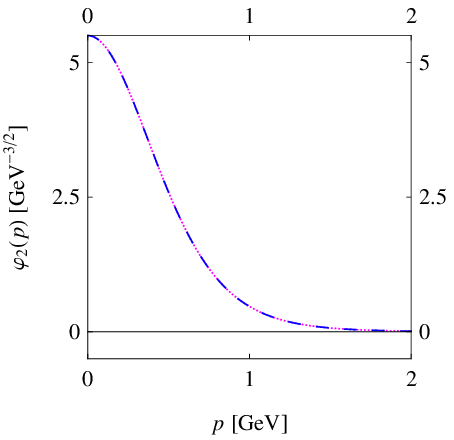}\\(a)&(b)
\end{tabular}\caption{Nonvanishing radial Salpeter component
function (\textcolor{blue}{dashed}) in configuration (a) and
momentum (b) space \cite{WL-N}, perfectly matching the $\widehat
M=0$ solution (\textcolor{magenta}{dotted}) to our
three-dimensional bound-state equation (\protect\ref{GIBSE}) with
the inversion-rooted potential $V(r)$ depicted in
Fig.~\protect\ref{Fig:PoV} \cite{WL05}, solved by application of
variational techniques (a) or conversion to an equivalent matrix
eigenvalue problem (b). Here, $p$ indicates the radial~variable
$p\equiv|\mathbf{p}|$.}\label{Fig:RSC}\end{center}\end{figure}

\begin{figure}[hbt]\begin{center}
\includegraphics[scale=1.6266]{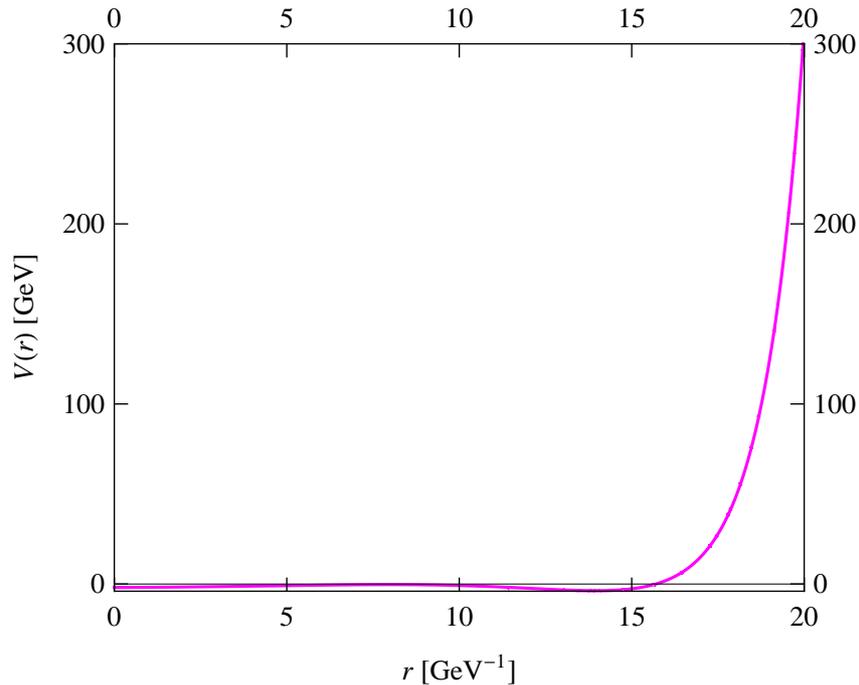}\caption{Potential
$V(r)$ \cite{WL-N} defining Fierz-symmetric effective interactions
providing a proper description of Goldstone-type pseudoscalar
mesons by the (three-dimensional) bound-state equation formulated
in Ref.~\cite{WL05}: rising in a confining manner from a finite
value, $V(0)=-1.92\;\mbox{GeV},$ after passing a zero at
$r=15.7\;\mbox{GeV}^{-1}$ to infinity, it resembles, in contrast
to earlier findings \cite{WL-I} focusing on particular aspects, a
smoothed square~well.}\label{Fig:PoV}\end{center}\end{figure}

In the exact Goldstone limit $\widehat M=0$,
immediately ensuring $\varphi_1(|\mathbf{p}|)=0$, the set
(\ref{GIBSE}) collapses to a single integral equation for the sole
surviving component $\varphi_2(|\mathbf{p}|)\not\equiv0$. From the
Fourier transform of that latter relation, the sought
configuration-space potential $V(r)$ is easily read off. We
extract our inversion input from a popular chiral-limit model
solution \cite{MT} to the QCD equation of motion for the full
quark propagator. Figure~\ref{Fig:RSC} illustrates our finding for
$\varphi_2$ in configuration and momentum~space.\footnote{The
predicted meson size makes sense \cite{WL-N}: average interquark
distance $\langle r\rangle=0.483\;\mbox{fm}$ and
root-mean-square~radius $\sqrt{\langle
r^2\rangle}=0.535\;\mbox{fm}$ nicely match the measured
electromagnetic charge radius $\sqrt{\langle
r_\pi^2\rangle}=(0.672\pm0.008)\;\mbox{fm}$ \cite{PDG} of
the~pion.} Using this in the Fourier transform of
Eq.~(\ref{GIBSE}), the behaviour of $V(r)$ given in
Fig.~\ref{Fig:PoV} can be~derived, regrettably only in numerical
form, but, at least within the range of $r$ depicted in
Fig.~\ref{Fig:PoV}, can be~easily represented in terms of
elementary functions, \emph{e.g.\/}, with the parameter values
collected in Table~\ref{Tab:M},~by\begin{equation}V(r)\approx
a_0+a_1\,r+a_2\,r^2+b\exp(c\,r)\ .\label{VE}\end{equation}

\begin{table}[h]\caption{Numerical values of the parameters
in our approximation (\protect\ref{VE}) to the potential $V(r)$
shown in~Fig.~\protect\ref{Fig:PoV}.}\label{Tab:M}
\begin{center}\begin{tabular}{lccccc}\hline\hline\\[-1.885ex]
Parameter&\multicolumn{1}{c}{$a_0\left[\mbox{GeV}\right]$}
&\multicolumn{1}{c}{$a_1\left[\mbox{GeV}^2\right]$}
&\multicolumn{1}{c}{$a_2\left[\mbox{GeV}^3\right]$}
&\multicolumn{1}{c}{$b\left[\mbox{GeV}\right]$}
&\multicolumn{1}{c}{$c\left[\mbox{GeV}\right]$}\\[.956ex]\hline
\\[-1.885ex]Value&$-3.229$&$1.0095$&$-0.083685$&$1.411\times10^{-5}$
&$0.8475$\\[.956ex]\hline\hline\end{tabular}\end{center}\end{table}

\end{document}